\documentclass[
    ,final            % use final for the camera ready runs
  ]
  {aipproc}

\layoutstyle{6x9}
\newcommand{\be}{\begin{equation}}
\newcommand{\ee}{\end{equation}}
\newcommand{\bea}{\begin{eqnarray}}
\newcommand{\eea}{\end{eqnarray}}

\newcommand{\g}{\gamma}
\newcommand{\f}{\frac}

\newcommand{\bra}{\langle}
\newcommand{\ket}{\rangle}
\newcommand\lr[1]{{\left({#1}\right)}}
\begin{document}

\title{Particle production and saturation at HERA}


\author{Cyrille Marquet}{
  address={Service de physique th{\'e}orique, CEA/Saclay,
  91191 Gif-sur-Yvette cedex, France\\
  URA 2306, unit{\'e} de recherche associ{\'e}e au CNRS}}

\begin{abstract}
Perturbative QCD in the high-energy limit describes the evolution of scattering 
amplitudes with increasing energy towards and into the so-called saturation 
regime. Comparisons of the predictions with experimental data for a number of 
observables led to significant progress and understanding. We discuss the case 
of particle-production cross-sections measured at HERA and argue that these 
measurements have the potential to provide evidence for the saturation regime of 
QCD.
\end{abstract}

\maketitle

%%%%%%%%%%%%%%%%%%%%%%%%%%%%%%%%%%%%%%%%%%%%
%% MAINMATTER
%%%%%%%%%%%%%%%%%%%%%%%%%%%%%%%%%%%%%%%%%%%%

\section{Introduction}

In the Regge limit of perturbative QCD, {\em i.e.} when the centre-of-mass 
energy in a collision is much bigger than the fixed hard scale of the 
problem, parton densities inside the colliding particules grow with increasing 
energy, leading to the growth of scattering amplitudes. When the parton 
density becomes too large and the scattering amplitudes approach the unitarity 
limit, one enters in a regime called saturation \cite{glr}. 

The transition to the saturation regime is characterized by the so-called
saturation scale which is an intrinsic hard scale of the problem. Contributions 
to the scattering amplitudes which are neglected as higher twist in the Bjorken 
limit of perturbative QCD become important in the saturation regime: 
leading-twist gluon distributions are no more sufficient to describe scattering 
at high energies.

A consistent approach is to express physical observables in terms of the leading 
terms in an expansion with respect to the inverse of the center-of-mass energy: 
this is called the eikonal approximation. This formalism is well-suited 
because, as the energy increases, density effects and non-linearities that lead 
to saturation and unitarization of the scattering amplitudes can be taken into 
account.

In the following, after introducing the eikonal approximation, we discuss the 
case of particle-production cross-sections measured at HERA. We concentrate on 
the phenomenology for diffractive observables: we review their model 
descriptions and investigate their potential to provide evidence for parton 
saturation.

\section{High-energy scattering and saturation}

\begin{figure}[t]
\includegraphics[width=12.5cm]{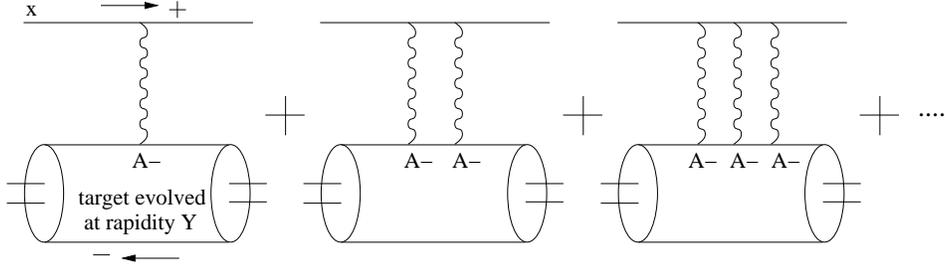}
\caption{Eikonal propagation of a quark with transverse position $x$ through a 
target evolved at rapidity $Y.$ The eikonal phase $W_F(x),$ see formula 
(\ref{wils}), resums all number of gluon exchanges.}
\label{eikonal}
\end{figure}

Let us start with the eikonal approximation for quarks and gluons scattering at 
high energies. When a system of partons propagating at nearly the speed of light 
passes through a target and interacts with its gluon fields, the dominant 
couplings are eikonal: the partons have frozen transverse coordinates and the 
gluon fields of the target do not vary during the interaction. This is justified 
since the time of propagation through the target is much shorter than the 
natural time scale on which the target fields vary. The effect of the 
interaction with the target is that the partonic components of the incident 
wavefunction pick up eikonal phases: if $|(\alpha,x)\ket$ (resp. $|(a,x)\ket$) 
is the wavefunction of an incoming quark of color index $\alpha\!\in\![1,N_c]$ 
(resp. gluon of color index $a\!\in\![1,N^2_c\!-\!1]$) and transverse position 
$x$ (the irrelevant degrees of freedom like spins or polarizations are not 
explicitly mentioned), then the action of the $S-$matrix is (see for 
instance \cite{eikap}):
\be
S|(\alpha,x)\ket\!\otimes\!|t\ket=
\sum_{\alpha'}\left[W_F(x)\right]_{\alpha\alpha'}
|(\alpha',x)\ket\!\otimes\!|t\ket\ , \hspace{0.3cm}
S|(a,x)\ket\!\otimes\!|t\ket=\sum_{b}W_A^{ab}(x)
|(b,x)\ket\!\otimes\!|t\ket\ ,\label{smat}\ee
where $|t\ket$ denotes the initial state of the target. The phase shifts due to 
the interaction are the color matrices $W_F$ and $W_A,$ the eikonal 
Wilson lines in the fundamental and adjoint representations respectively, 
corresponding to propagating quarks and gluons. They are given by
\begin{equation}
W_{F,A}(x)=P\exp\{ig_s\int dz_+T_{F,A}^aA_-^a(x,z_+)\}
\label{wils}\end{equation}
with $A_-$ the gauge field of the target and $T_{F,A}^a$ the 
generators of $SU(N_c)$ in the fundamental ($F$) or adjoint ($A$)
representations. We use the light-cone gauge $A_+\!=\!0$ and
$P$ denotes an ordering in the light-cone variable $z_+$ 
along which the incoming partons are propagating. As displayed in Fig.1, all 
number of gluons exchanges are included in (\ref{smat}), which shows the leading 
term of an expansion with respect to the inverse of the center-of-mass energy 
squared $s\!\sim\!e^{Y}$ where $Y$ is the total rapidity.

For an incoming state $|\Psi_{in}\ket,$ the outgoing state 
$|\Psi_{out}\ket\!=\!S|\Psi_{in}\ket\!\otimes\!|t\ket$ emerging from the 
eikonal interaction is obtained by the action of the 
$S-$matrix on the partonic components of $|\Psi_{in}\ket$ as indicated by 
formula (\ref{smat}). The outgoing wavefunction $|\Psi_{out}\ket$ is therefore a 
function of the Wilson lines (\ref{wils}). When calculating physical observables 
from $|\Psi_{out}\ket$, one obtains objects that are target averages of traces 
of Wilson lines (the traces come from the color summation). For instance, the 
simplest of these objects is
\be 
T_{q\bar q}(x,x';Y)=1-
\f1{N_c}\left\bra\mbox{Tr}\lr{W_F^\dagger(x')W_F(x)}\right\ket_Y\ , 
\label{qqdip}\ee
namely the $q\bar q-$dipole scattering amplitude ($x,$ $x':$ positions of the 
quark and antiquark) off a target evolved at rapidity $Y$. The target average 
has been denoted $\bra\ .\ \ket_Y$ and contains the $Y$ 
dependence. The amplitude (\ref{qqdip}) enters for instance in the DIS total 
cross-section (see next section). More generally, observables are functions of 
(\ref{qqdip}) or more complicated amplitudes. Let us introduce another one of 
them, which we shall need later:
\be 
T^{(2)}_{q\bar q}(x,x';y,y';Y)=1-
\f1{N_c^2}\left\bra\mbox{Tr}\lr{W_F^\dagger(x')W_F(x)}
\mbox{Tr}\lr{W_F^\dagger(y')W_F(y)}\right\ket_Y\  
\label{2qqdip}\ .\ee
This is the scattering amplitude for a set of two dipoles ($x,x'$) and ($y,y'$). 
The amplitudes (\ref{qqdip}) and (\ref{2qqdip}) take values between 0 
(transparency) and the black-disk (saturation) limit 1.

To actually compute these amplitudes, one has to evaluate the averages $\bra\ .\ 
\ket_Y$ which amounts to calculating averages of Wilson lines: $\left\bra f[A] 
\right\ket_Y\!=\! \int {\cal D}A f[A] U_Y[A]$ where the target wavefunction 
$U_Y[A]$ 
represents the probability to find a given field configuration inside the target 
evolved at rapidity $Y.$ The information contained in the target averages is 
mainly non-perturbative but the evolution towards higher rapidities $dU_Y[A]/dY$ 
can be computed perturbatively, at least in the leading-logarithmic 
approximation. Several equations have been established with different degrees of 
approximations, we shall not discuss them here and the reader can refer to 
\cite{proc} for more details. Let us only mention the Balitsky-Kovchegov 
saturation equation (BK) \cite{bk} which is a closed equation for $T_{q\bar q}$ 
obtained in a mean-field approximation. We shall refer to the BK equation later 
on when we link observables to the dipole amplitudes (\ref{qqdip}) and 
(\ref{2qqdip}) and discuss phenomenology.

\section{Saturation phenomenology at HERA}

In deep inelastic scattering (DIS), a photon of virtuality $Q^2$ collides
with a proton. In an appropriate ``dipole'' frame, the virtual photon undergoes 
the hadronic interaction via a fluctuation into a dipole (see Fig.2, left); the 
dipole then interacts with the target proton and one has the following 
factorization
\be
\sigma_{DIS}(Q^2,Y)=\int d^2r\ \psi(|r|,Q^2)\ 
2\int d^2b\ T_{q\bar q}\lr{b-\f r2,b+\f r2;Y}
\ee
which relates the DIS total cross-section $\sigma_{DIS}$ to the $q\bar q-$dipole 
amplitude $T_{q\bar q}.$ The function 
$\psi(r,Q^2)\!=\!\int dz\ |\phi^{\g}(r,z;Q^2)|^2$ is obtained from the 
well-known wavefunction $\phi^{\g}(r,z;Q^2)$ which describes the splitting of 
the photon 
onto a dipole of transverse size $r$ and with the antiquark carrying a fraction 
of photon longitudinal momentum $z.$ Note that in this case, not all the 
information on $T_{q\bar q}$ is relevant as the impact parameter $b$ is 
integrated out: only the cross-section $\sigma_{q\bar q}(r,Y)\!=\!2\int d^2b\ 
T_{q\bar q}\lr{b\!-\!\f r2,b\!+\!\f r2;Y}$ is needed.

\begin{figure}[t]
\hspace{0.8cm}
\begin{minipage}[t]{45mm}
\includegraphics[width=3.6cm]{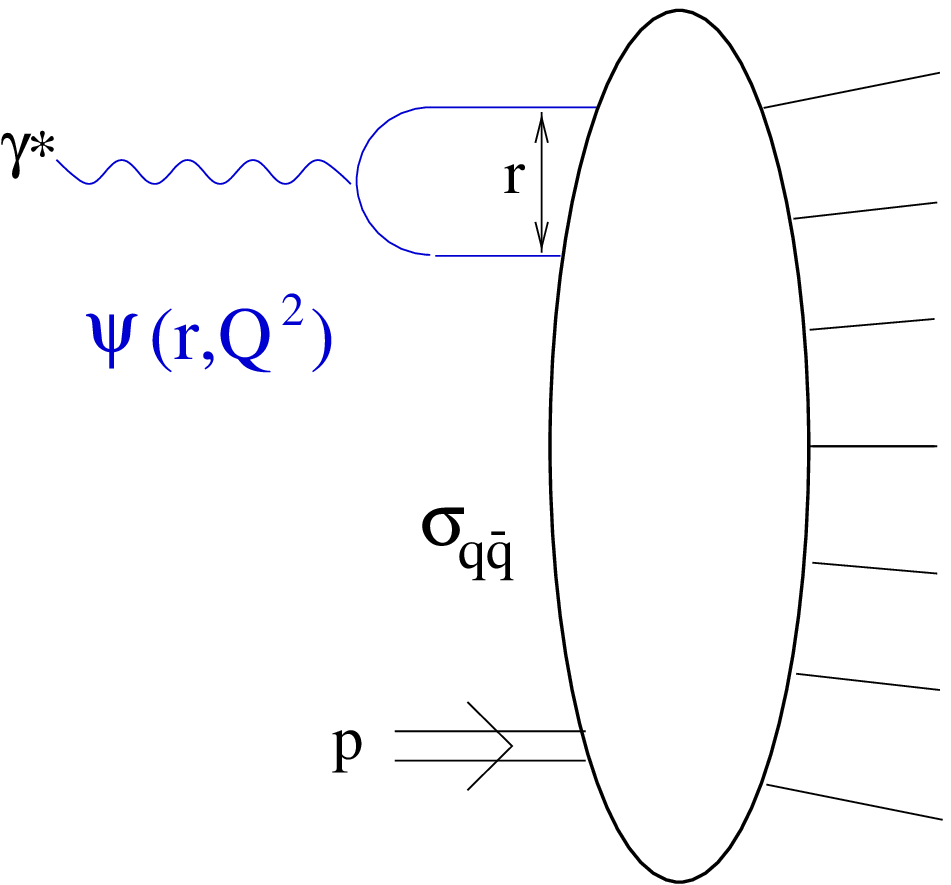}
\end{minipage}
\hspace{\fill}
\begin{minipage}[t]{60mm}
\includegraphics[width=5.3cm]{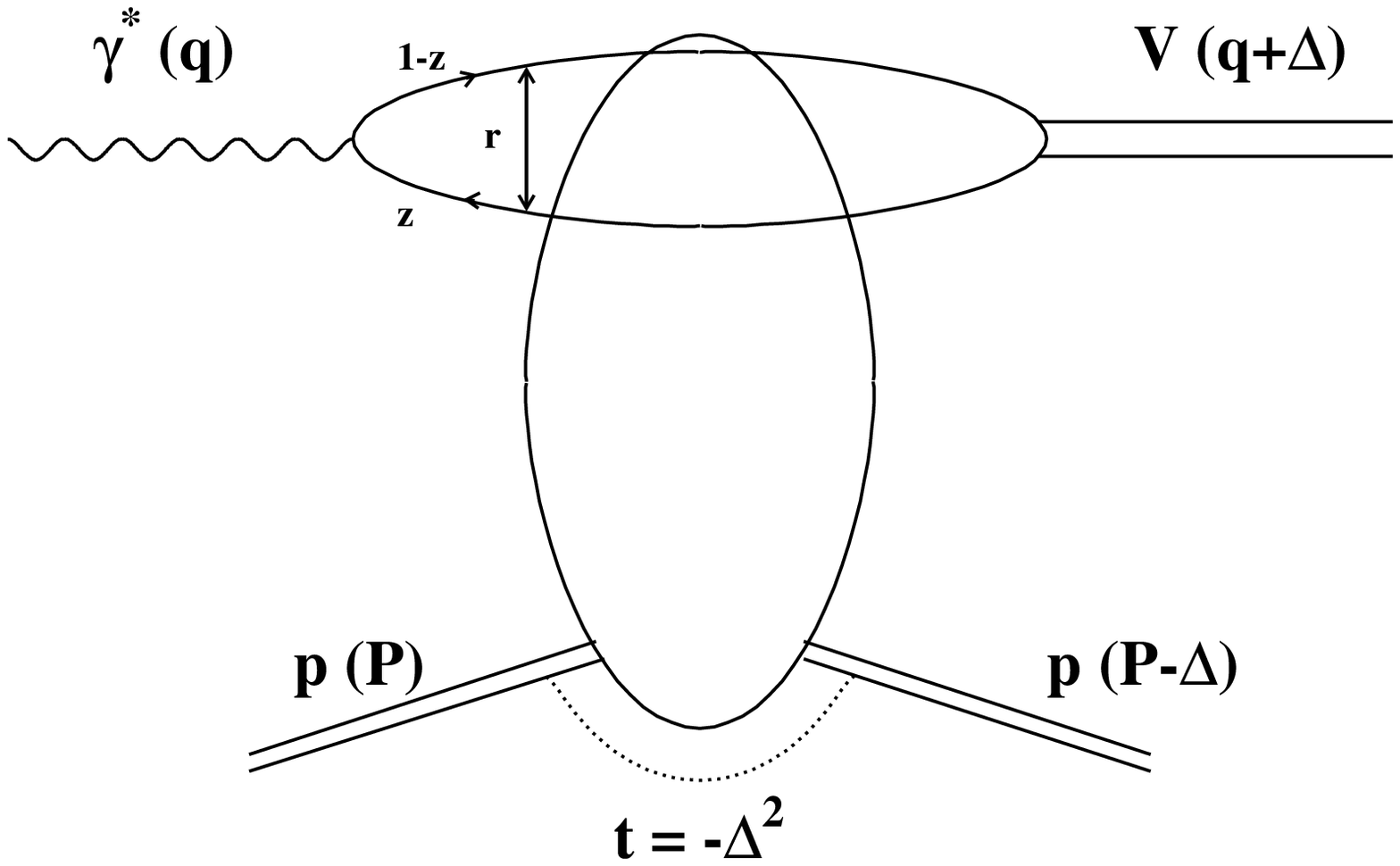}
\end{minipage}
\hspace{\fill}
\begin{minipage}[t]{45mm}
\includegraphics[width=3.1cm]{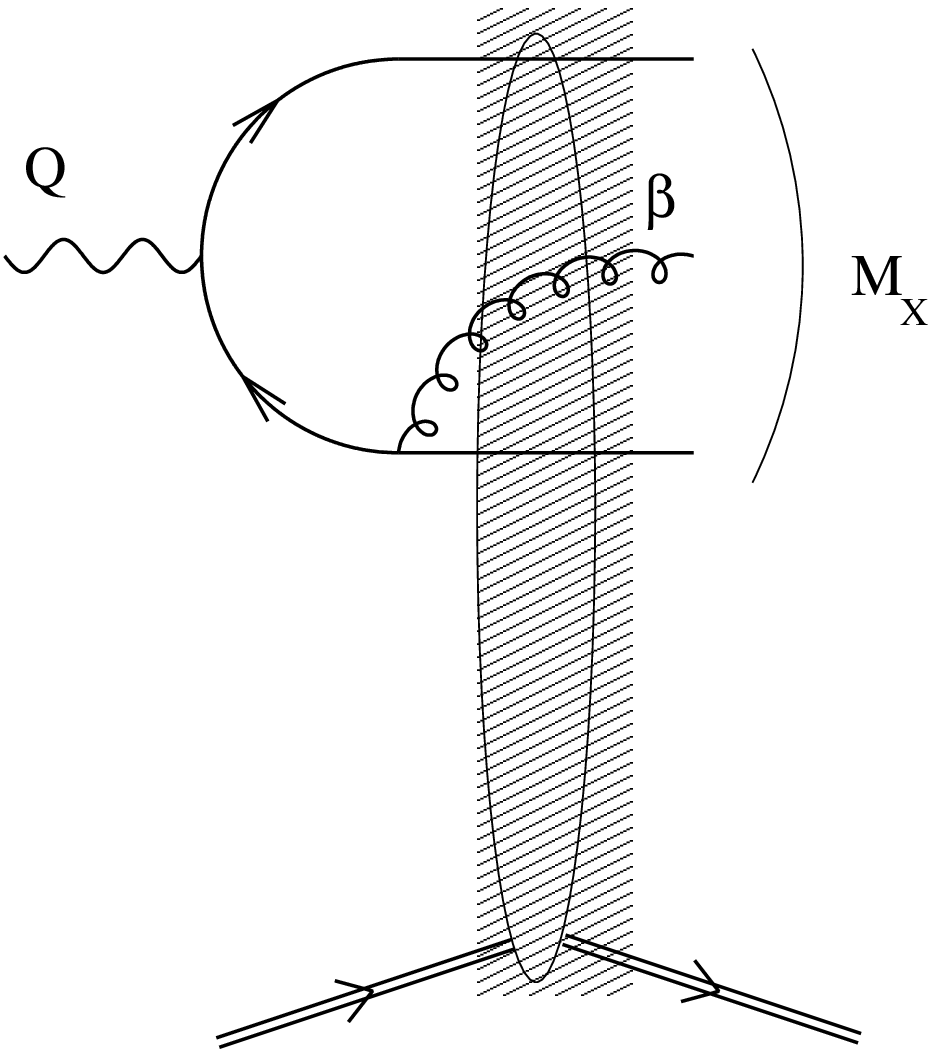}
\end{minipage}
\caption{Three processes measured in virtual photon-proton collisions at HERA: 
DIS total cross-section (left), diffractive vector-meson production (center), 
and diffractive photon dissociation (right).}
\label{dis}
\end{figure}

Measurements of $\sigma_{DIS}$ at HERA have had a great impact on saturation 
phenomenology, especially the discovery of geometric scaling \cite{geom}: the 
fact that $\sigma_{DIS}(Q^2,Y)$ is a function of the single variable 
$Q^2/Q_s^2(Y)$ with the saturation scale $Q_s^2(Y)\!\sim\!\exp{(\lambda Y)}$ and 
$\lambda\!\simeq\!0.28.$ Indeed this has 
a natural explanation in terms of traveling-wave solutions \cite{tv} of the BK 
equation $T_{q\bar q}(r,b\!=\!0,Y)\!=\!T_{q\bar q}(rQ_s(Y)).$ However, this 
result is obtained neglecting the impact parameter dependence of $T_{q\bar q}$ 
and considering $\sigma_{q\bar q}(r,Y)\!=\!S_P\!\times\! T_{q\bar 
q}(r,b\!=\!0,Y)$ where $S_P$ is the transverse area of the proton fitted to the 
data.

In order to understand better and study more consistently the impact parameter 
dependence of $T_{q\bar q},$ the authors of \cite{mms} have looked at 
diffractive vector-meson production (see Fig.2, center). In diffractive deep 
inelastic scattering, the proton gets out of the collision intact and there is a 
rapidity gap between that proton and the final state $X.$ When the final state 
is a vector meson, the momentum transfer $\Delta$ dependence of the 
cross-section is related to the impact parameter dependence of the dipole 
amplitude. Indeed the cross-section reads ($t\!=\!-\Delta^2$)
\be
\f{d\sigma}{dt}=\f1{4\pi}\left|\int d^2r\ \Psi(|r|,Q^2,M_V^2)\ 
\int d^2b\ e^{ib.\Delta}\ T_{q\bar q}\lr{b-\f r2,b+\f r2;Y}\right|^2
\label{tspec}\ee
where the function $\Psi(r,Q^2,M_V^2)\!=\!\int dz\ 
\phi^{\g}(r,z;Q^2)\phi^{V}(r,z;M_V^2)$ is obtained from both the photon 
wavefunction $\phi^{\g}$ and the final-state meson (whose 
mass has been denoted $M_V$) wavefunction $\phi^V.$ By analysing data on 
$\rho-$meson production at fixed $Q^2$ and $Y\!\simeq\!7$, the authors extracted 
the dipole $S-$matrix $S_{q\bar q}(r,b;Y)\!=\!1\!-\!T_{q\bar q}\lr{b\!-\!\f 
r2,b\!+\!\f r2;Y}$ as a function of $b$ for a fixed size $r_Q$ with 
$r_Q^2\!\sim\!4/(Q^2\!+\!M_V^2).$ Three different sets of data at different 
$Q^2$ have been used. Their results are shown on the left plot of Fig.3; the 
shaded area on the left is an uncontrolled region due to the lack of large$-t$ 
data. The plot shows that the $b$ dependence cannot be neglected and that 
$T_{q\bar q}(r\!\sim\!1\ \mbox{GeV}^{-1},b\!\sim\!0;Y\!\sim\!7)\!\simeq\!0.4.$ 
This significant value of $T_{q\bar q}$ indicates that HERA could be entering 
the saturation regime.

As the importance of the impact parameter had been pointed out, a 
phenomenological model for the dipole amplitude $T_{q\bar q}$ with an impact 
parameter profile was proposed in \cite{kt}. With that saturation  
parametrization, the authors could well reproduce the data for diffractive 
$J/\Psi$ production at HERA: the $t$ spectrum (\ref{tspec}) as well the the 
$Q^2$ and $Y$ dependences of the total cross-section $\sigma_{J/\Psi}.$ Their 
results are displayed on the center and right plots of Fig.3 where one can see 
the good agreement of the model with the data. A successful description of the 
same data using numerical simulations of the BK equation was also given in 
\cite{gllmn}, confirming the compatibility of saturation predictions.

However in all the model descriptions of $t$ spectra, the impact parameter 
dependence was introduced by hand as one had not extracted any information on 
the $b$ depencence of $T_{q\bar q}$ from saturation equations. That moderated 
the impact of the results mentioned above. Interestingly, it was recently 
\cite{tvmom,bkmom} pointed out that important information can be obtained from 
the BK equation when looking at the $\Delta$ dependence of 
$\tilde{T}(r,\Delta;Y)\!=\!\int d^2b\ e^{ib.\Delta}\ T_{q\bar q}\lr{b\!-\!\f 
r2,b\!+\!\f r2;Y}.$ For instance, the geometric scaling property was extended: 
$\tilde{T}(r,\Delta;Y)\!=\!\tilde{T}(rQ_s(\Delta,Y))$ with 
$Q^2_s(\Delta,Y)\!\sim\!\Delta^2\exp{(\lambda Y)}.$ Parametrizing the dipole 
amplitude with the momentum transfer instead of the impact parameter opens a new 
approach to analyse the data. An experimental confirmation of geometric scaling 
at non-zero momentum transfer would represent a significant success for
saturation.

\begin{figure}[t]
\begin{minipage}[t]{60mm}
\includegraphics[width=5.3cm]{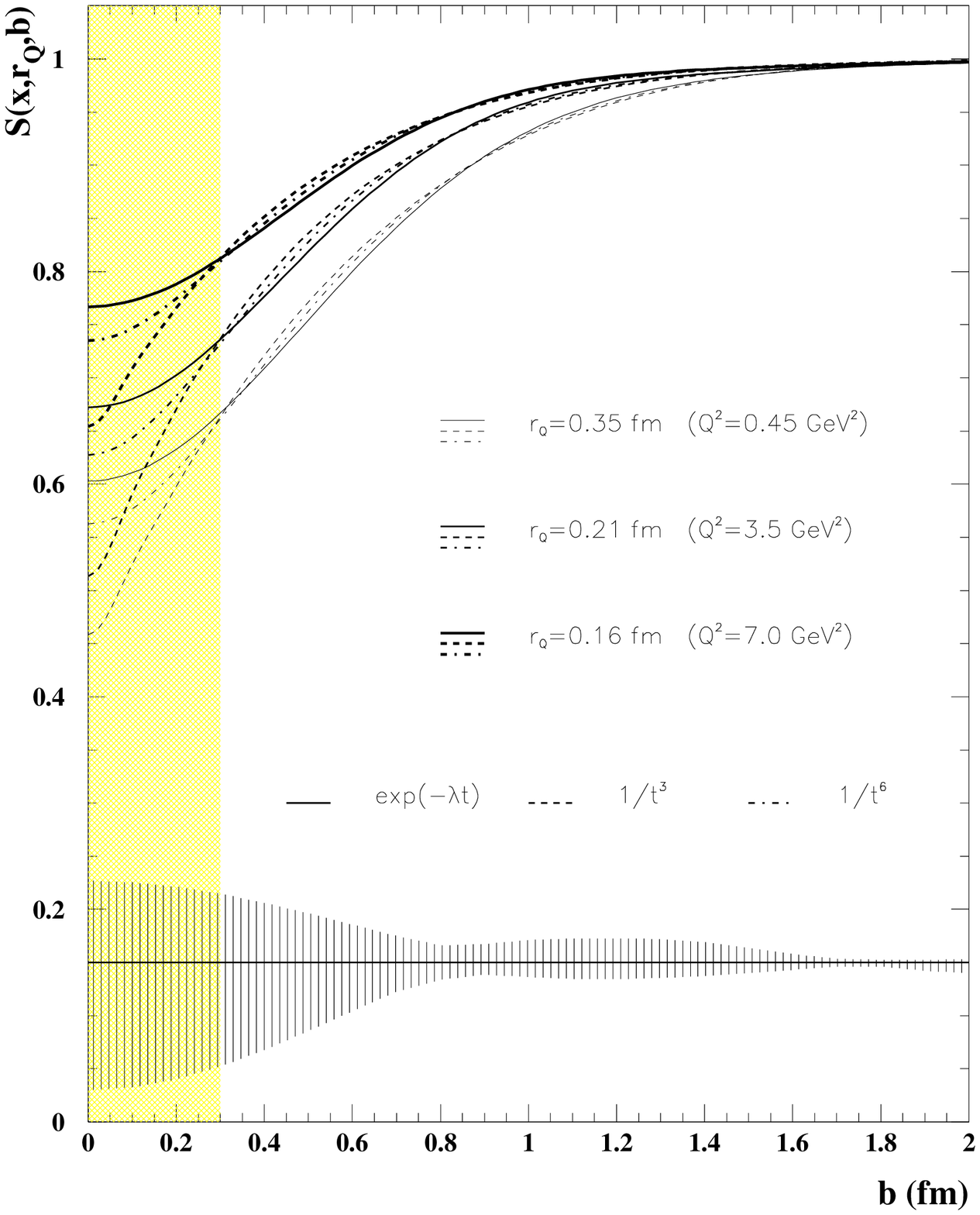}
\end{minipage}
\hspace{\fill}
\begin{minipage}[t]{50mm}
\includegraphics[width=4.6cm]{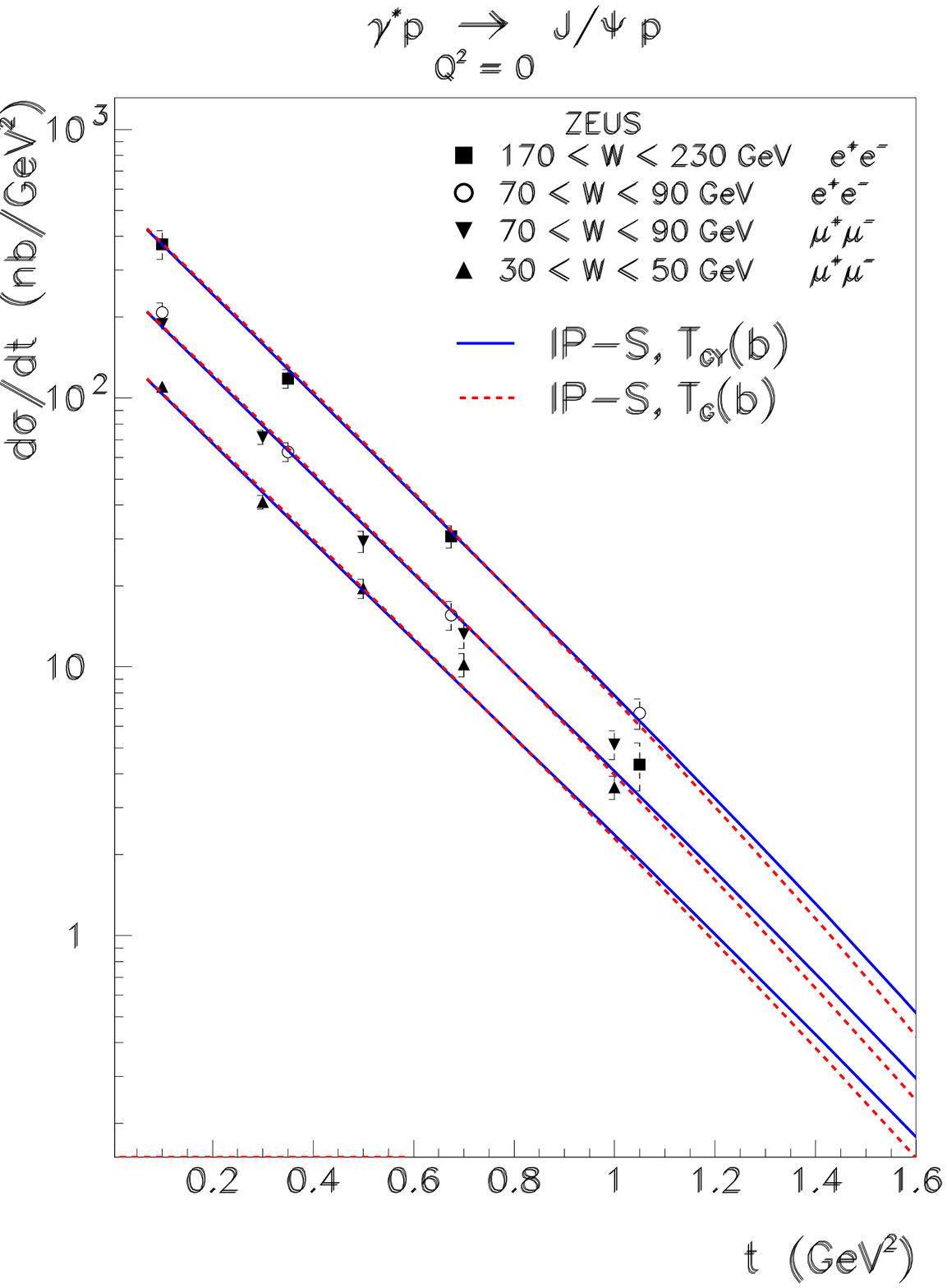}
\end{minipage}
\hspace{\fill}
\begin{minipage}[t]{40mm}
\vspace{-6.5cm}\includegraphics[width=3.5cm]{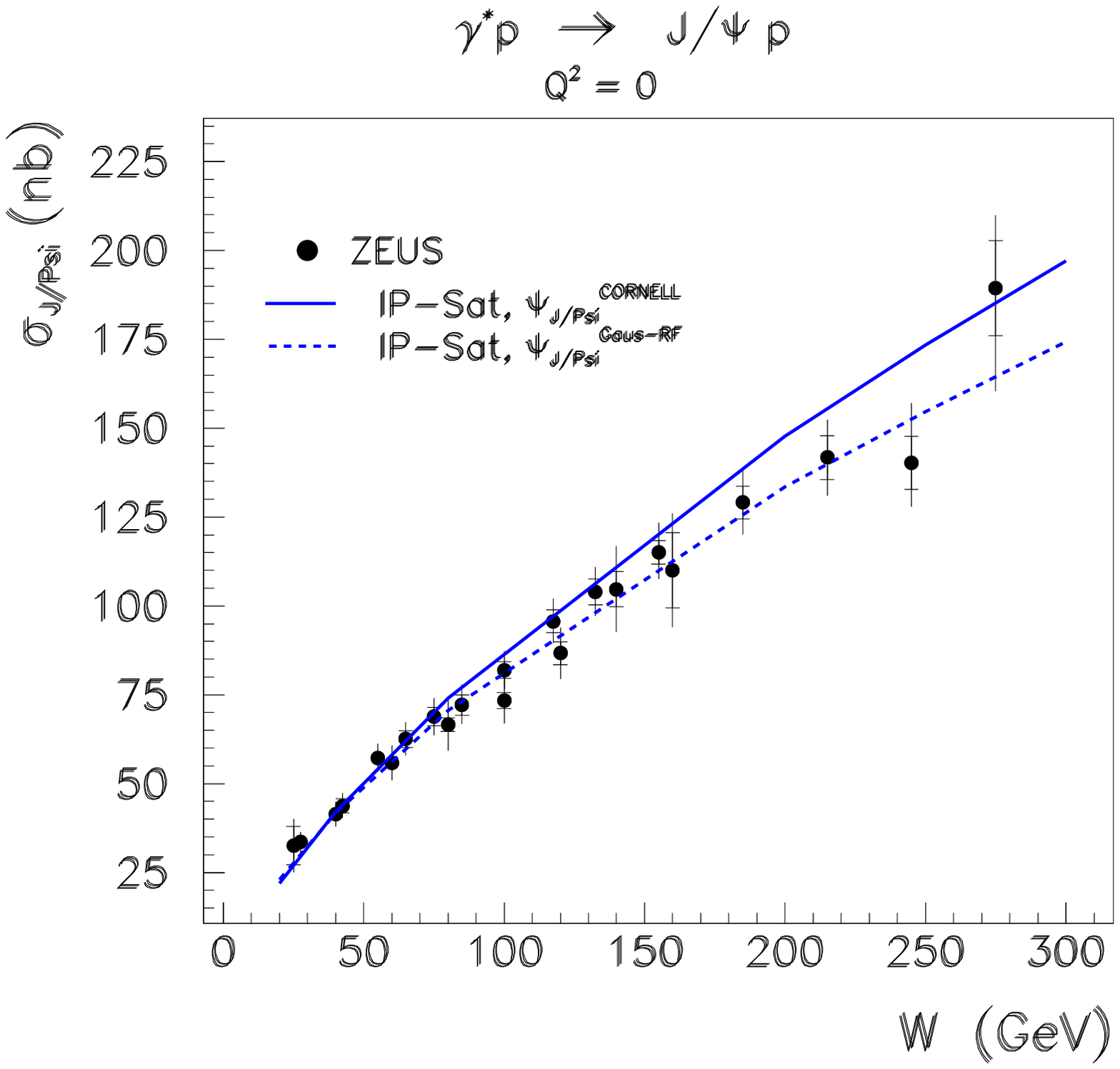}
\vfill\includegraphics[width=3.5cm]{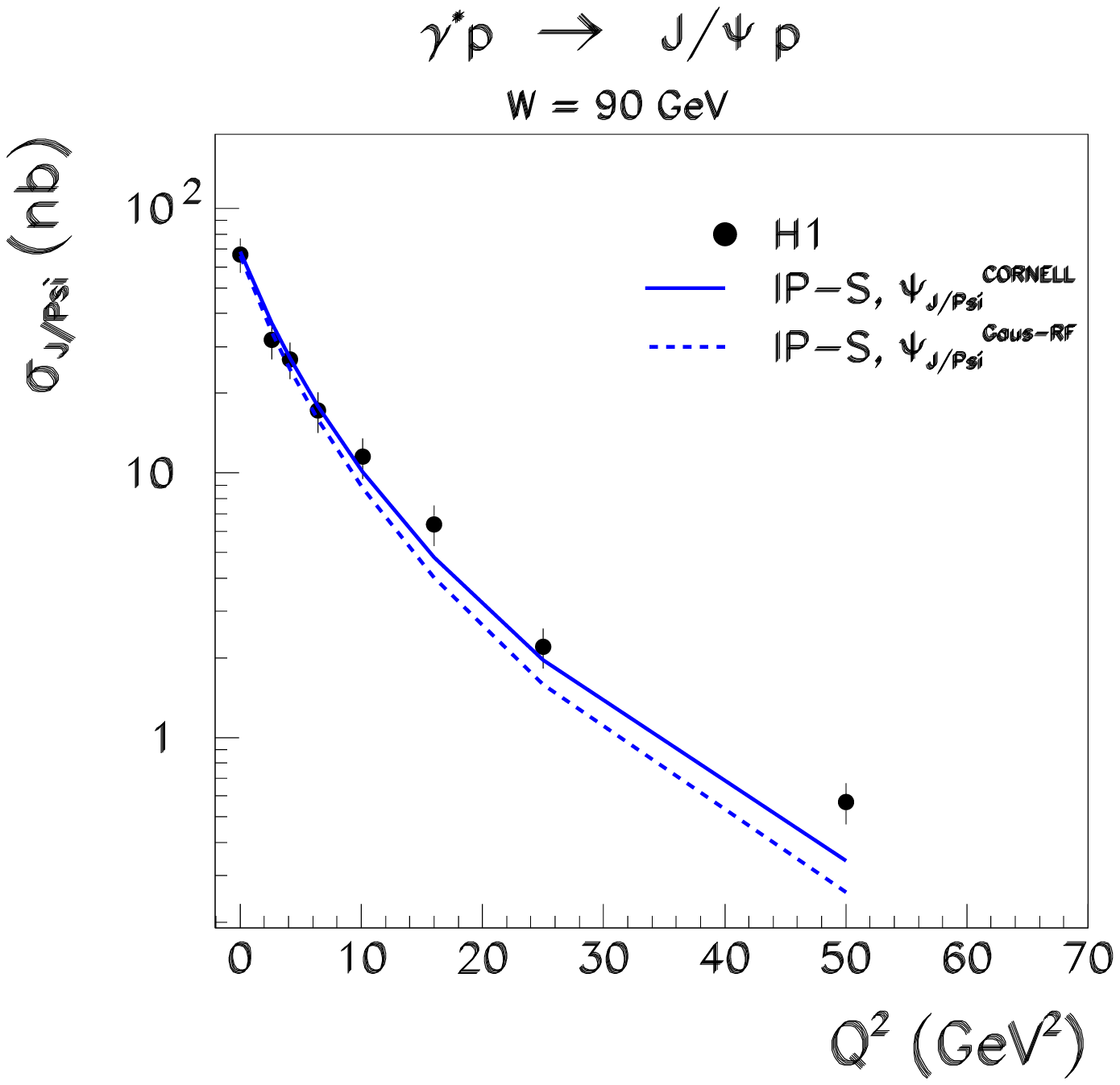}
\end{minipage}
\caption{Left plot: the $q\bar q-$dipole $S-$matrix extracted from the 
diffractive 
$\rho-$meson production data in \cite{mms}; this shows the impact parameter 
dependence for three different dipole sizes $r_Q$ and a rapidity $Y\!\simeq\!7.$
Center and right plots: diffractive $J/\Psi$ production at HERA; $d\sigma/dt$ as 
a function of $t$ (center), $\sigma_{J/\Psi}$ as a function of 
$W\!\sim\!\exp({Y/2})$ (top right) and $Q^2$ (bottom right); comparison with the 
model of \cite{kt} is shown.}
\label{kt}
\end{figure}

Let us finally consider high-mass diffraction. If the final state diffractive 
mass $M_X$ is much bigger than $Q,$ the dominant configurations to the 
final state come from the $q\bar q g$ component of the photon wavefunction (see 
Fig.2, right) or from higher Fock states, i.e. from the photon dissociation. By 
contrast, if $M_X\!\ll\!Q,$ the dominant configurations come from the $q\bar q$ 
component as it was the case for vector-meson production. Let us then consider 
the kinematical regime where $\beta\!\equiv\!Q^2/(Q^2\!+\!M_X^2)\!\ll\!1$ and 
investigate the $q\bar q g$ component. The right plot of Fig.2 represents the 
diffractive production of a gluon with transverse momentum $k$ and rapidity 
$\log(1/\beta)$ in the collision of the photon with the target proton. Provided 
$k$ is a hard scale, the gluon momentum spectrum is given by~\cite{marq}
\be
\label{sigd}
\f{M_X d\sigma}{d^2kdM_X}=\f{\alpha_sN_c^2}{2\pi^2C_F}
\int d^2r\ \psi(|r|,Q^2) \int d^2b\ {\bf A}(k,r\!-\!\f b2,r\!+\!\f 
b2;\Delta\eta)\cdot 
{\bf A}^*(k,r\!-\!\f b2,r\!+\!\f b2;\Delta\eta)
\ee
where $\Delta\eta\!=\!Y\!-\!\log(1/\beta)$ is the rapidity gap. The 
two-dimensional vector ${\bf A}$ is given by
\be
{\bf A}(k,x,x';\Delta\eta)=\int\f{d^2z}{2\pi}\ e^{-ik.z}
\left[\f{z\!-\!x}{|z\!-\!x|^2}-\f{z\!-\!x'}{|z\!-\!x'|^2}\right]
\lr{T^{(2)}_{q\bar q}(x,z;z,x';\Delta\eta)-T_{q\bar q}(x,x';\Delta\eta)}
\label{ampla}.
\ee

Interestingly enough, independently of the form of the dipole amplitudes 
$T_{q\bar q}$ and $T^{(2)}_{q\bar q},$ the behavior of the observable $k^2\ 
d\sigma/d^2kdM_X$ as a function of the gluon transverse momentum $k$ is the 
following \cite{golmar}: it rises as $k^2$ for small values of $k$ and falls as 
$1/k^2$ for large values of $k.$ A maximum occurs for a value $k_0$ which is 
related to the inverse of the typical size for which the $T-$matrices approach 
one; in other words, the maximum $k_0$ reflects the scale at which unitarity 
sets in. If the energy is large enough so that the saturation scale $Q_s$ is 
hard, unitarity will come as a consequence of parton saturation and 
$k_0\!\sim\!Q_s.$ If not the case, unitarity will be rather due to 
non-perturbative physics.

In the saturation case, the model of \cite{golmar} for the dipole amplitudes 
allows to plot the whole $k$ spectrum (\ref{sigd}). This is shown on the left 
plot of Fig.4 and one can indeed see that the spectrum features a maximum peaked 
around $k_0\!\simeq\!1.4\ Q_s$ independently of $Q^2$ and $Q_s.$ Measuring this 
cross-section at HERA would offer a unique opportunity to test if 
saturation plays a role in diffraction at the present energies. On the right 
plot of Fig.4, the cross-section is plotted in the HERA energy range for 
different values of $M_X$ and total energy $W\!\sim\!e^{Y/2},$ corresponding to 
different values of $Q_s.$ The saturation scale is the one extracted 
\cite{golec} from the $F_2$ data. As expected for realitic jet transverse 
momenta, $k\!>\!Q_s$ and the data would lie on the perturbative side of the 
bump. There is a big difference in the rise towards the bump between the lowest 
(top right) and highest (bottom left) $Q_s$ bins. A confirmation of this 
behavior would certainly favor the saturation scenario.

\begin{figure}[t]
\begin{minipage}[t]{70mm}
\includegraphics[width=6.3cm]{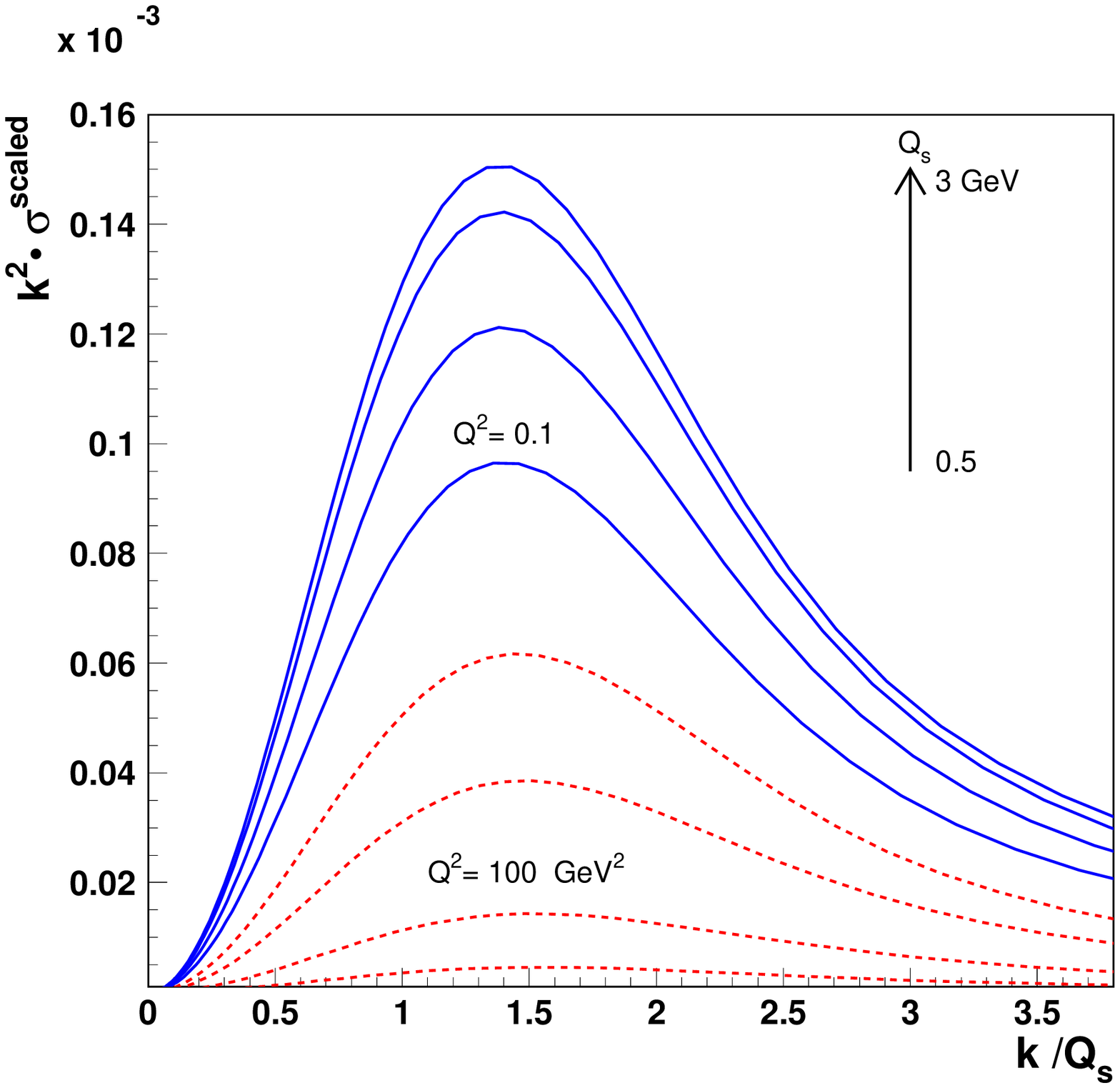}
\end{minipage}
\hspace{\fill}
\begin{minipage}[t]{70mm}
\hspace{1cm}\includegraphics[width=5.8cm]{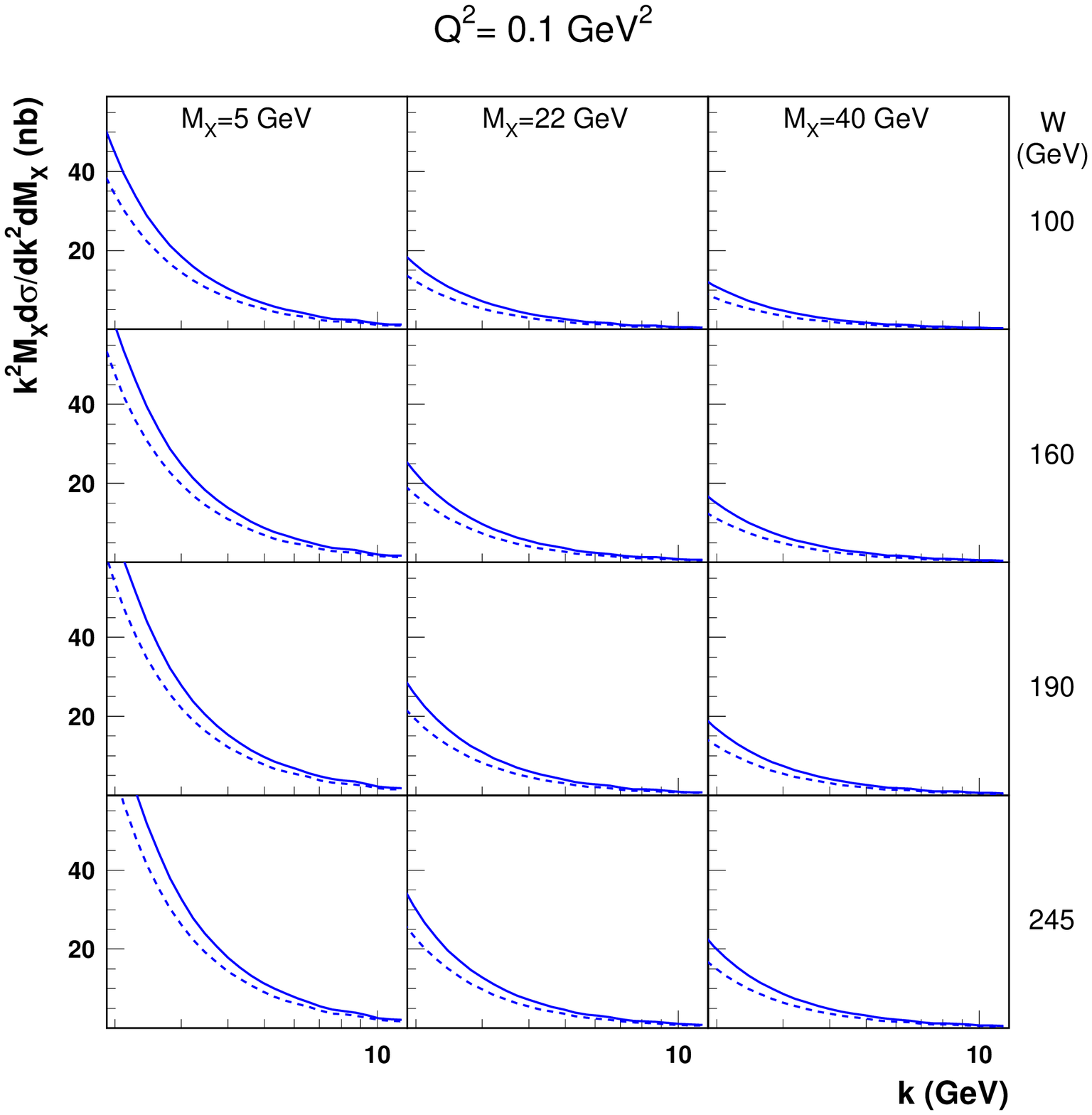}
\end{minipage}
\caption{The diffractive gluon production cross-section 
$k^2M_Xd\sigma/dk^2dM_X.$ Left plot: as a function of the rescaled gluon 
transverse momentum $k/Q_s$ for two extreme values of $Q^2$ equal to $0.1$ and 
$100~{\rm GeV}^2$ and four values of the saturation scale 
$Q_s\!=\!0.5,1,2,3~{\rm GeV}.$
Right plot: as a function of the jet transverse momentum $k$ and in the HERA 
energy range for $Q^2\!=\!0.1~{\rm GeV}^2$ and different values of diffractive 
mass $M_X$ and energy $W$; full lines: only the light quarks are included in 
$\psi$, dashed lines: charm is also included.}
\label{dgp}
\end{figure}

%%%%%%%%%%%%%%%%%%%%%%%%%%%%%%%%%%%%%%%%%%%%%%%%
%% BACKMATTER
%%%%%%%%%%%%%%%%%%%%%%%%%%%%%%%%%%%%%%%%%%%%%%%%

%%\begin{theacknowledgments}
 
%%\end{theacknowledgments}



\end{document}